\documentclass[runningheads]{llncs}
\usepackage[utf8]{inputenc}
\usepackage{longtable}
\usepackage{supertabular,booktabs}
\usepackage{caption}
\usepackage{multirow}
\usepackage{graphicx}
\usepackage{float}
\usepackage{comment}

\begin{document}

\title{SherLOCKED: A Detective-themed Serious Game for Cyber Security Education \thanks{This is the author version of the article: Jaffray A., Finn C., Nurse J.R.C. (2021) SherLOCKED: A Detective-Themed Serious Game for Cyber Security Education. In: Furnell S., Clarke N. (eds) Human Aspects of Information Security and Assurance. HAISA 2021. IFIP Advances in Information and Communication Technology, vol 613. Springer, Cham. https://doi.org/10.1007/978-3-030-81111-2{\_}4}}

\titlerunning{SherLOCKED: A Serious Game for Security Education}

\author{Alice Jaffray \and
Conor Finn \and
Jason R.C. Nurse
}
\authorrunning{A. Jaffray et al.}
%
\institute{University of Kent, UK\\ 
\email{J.R.C.Nurse@kent.ac.uk}
}

\maketitle              

\begin{abstract}
Gamification and Serious Games are progressively being used over a host of fields, particularly to support education. Such games provide a new way to engage students with content and can complement more traditional approaches to learning. This article proposes SherLOCKED, a new serious game created in the style of a 2D top-down puzzle adventure. The game is situated in the context of an undergraduate cyber security course, and is used to consolidate students' knowledge of foundational security concepts (e.g. the CIA triad, security threats and attacks and risk management). SherLOCKED was built based on a review of existing serious games and a study of common gamification principles. It was subsequently implemented within an undergraduate course, and evaluated with 112 students. We found the game to be an effective, attractive and fun solution for allowing further engagement with content that students were introduced to during lectures. This research lends additional evidence to the use of serious games in supporting learning about cyber security. 

\keywords{Cyber security \and Gamification \and Serious Games \and Education \and University \and Puzzle \and Detective \and COVID-19 \and Emergency Online Learning }
\end{abstract}

\section{Introduction}
Serious games are a unique opportunity for educators as they provide another, potentially more appealing, way to engage students with course content. The concept of a serious game has existed for decades and while many definitions exist, these games can be regarded as those with ``an explicit and carefully thought-out educational purpose and are not intended to be played primarily for amusement''~\cite{abt1987serious}. A salient point about serious games is that although amusement is not a primary objective, striking a good balance between being entertaining and educational can increase the game's appeal in learning settings. This appeal may allow educators to reach students who may be less interested in traditional forms of teaching and revision. 

There have been numerous research efforts exploring the development and use of games to educate individuals about cyber security. These have concentrated on game designs including table-top exercises, tower defence games, role playing games, simulations and puzzles~\cite{hart2020riskio,jin2018evaluation,mostafa2019development}. These approaches have had varying levels of success and impact in educational and awareness settings.  

This paper seeks to complement current research into gamification for cyber security education through the proposal and evaluation of a new serious game, namely SherLOCKED, targeted at undergraduate computer science university students. SherLOCKED is created in the style of a 2D top-down puzzle adventure meant to support further engagement with content delivered via lectures. We draw our primary novelty from two areas. 

The first is contextual and is grounded in the fact that in-person learning has been drastically reduced due to the COVID-19 pandemic. Educators and students have had to switch suddenly to online learning (also termed, emergency online learning~\cite{aguilera2020college}), a reality that has impacted student motivation and cognitive engagement~\cite{aguilera2020college}. We sought to support students during this difficult time, and therefore developed and deployed SherLOCKED on an undergraduate course. This paper is therefore the first to our knowledge to engage in a trial and evaluation of a serious game in such a unique context. 

The second key contribution of our work is in the reasonably large sample size for such an evaluation. Serious games for teaching security explored in current literature~\cite{deeb2019teaching,hart2020riskio,mostafa2019development,scholefield2019gamification,schreuders2016gamification,vsvabensky2018enhancing} often interact with small numbers of university students (e.g. N$<$30) which impacts the inferences that can be made. SherLOCKED is evaluated through a survey of 112 university students, where we collect feedback through quantitative and qualitative data. This therefore acts as one of the largest studies to date examining the perception and use of a serious game within an undergraduate cyber security cohort. 

The remainder of this article is structured as follows. Section~\ref{sec:lit} reviews previous salient attempts at gamification. In Section~\ref{sec:game}, we introduce SherLOCKED and present the principles guiding its creation. The user study is outlined and discussed in Section~\ref{sec:eval}, before concluding the report in Section~\ref{sec:con}.

\section{Literature Review}
\label{sec:lit}
Gamification has been explored to various extents in prior work. Serious games for general security awareness are arguably the most popular. Anti-phishing Phil~\cite{sheng2007anti} for instance is one of the most well-known games that has sought to educate people about detecting phishing attacks. The domain of phishing attracts a large amount of gamification research~\cite{roepke2020pond}, likely due to the prominence of phishing and its perception as a user-oriented threat. Beyond phishing, topics such as password security and cryptography also feature. Sholefield and Shepherd~\cite{scholefield2019gamification} design a role-playing quiz application (RPG) to educate the general population about good password practices. Their evaluation highlights the importance of games as an enjoyable way to learn, but also the difficulties in such pursuits (e.g. challenges in implementing effective leader boards). Similar positive findings are found by Deeb and Hickey~\cite{deeb2019teaching} as they explore the use of a 3D escape room game to teach students about cryptography.

Offline serious games present another way to engage individuals. Riskio is a tabletop game to raise awareness of cyber security concepts for those in business and for those studying security at university~\cite{hart2020riskio}. It is oriented around playing the roles of attackers and defenders within an organisational security context. Crypto Go is another physical card game proposed which can be used for educating about security, particularly cryptography~\cite{gonzalez2020shuffle}. Through user workshops, researchers found that the game improved motivation to study the topic and the understanding of the field. 

Focusing specifically on formal teaching contexts, Jin et al.~\cite{jin2018evaluation} situate their research on the growing need for a security workforce and use games to educate high school students. They propose and evaluate four cyber security education games (e.g. using virtual reality and tower defence) to teach topics such as security foundations, secure online behaviour, cyber-attack and defense methods and social engineering. Results were highly positive, and games were favoured by students and staff. Mostafa et al.~\cite{mostafa2019development} also explore multiple games for teaching security through their testing of six games and how well they were received by university students. The games spanned topics such as network attacks, key management and web security, were implemented as image puzzles, simulations, role playing and action/adventure genres. Based on a user study, they conclude that the games could contribute greatly to the educational process. 

Lastly, capture the flag (CTF) games and exercises are extremely popular in cyber security. They allow participants (many of which may be students new to the field) to learn about the technical aspects of security, including finding and exploiting vulnerabilities (thus capturing `flags'). {\v{S}}v{\'a}bensk{\`y} et al. \cite{vsvabensky2021cybersecurity} provide a recent overview of the field and highlight the various types of challenges implemented to teach security. A key finding of their work is that while CTFs clearly are an attractive proposition alongside traditional lectures, they currently predominately focus on technical knowledge but often neglect the human aspects of security; this is clearly a shortcoming given how much cybercriminals use these factors~\cite{nursecybercrime}. More specifically, we have seen CTFs applied for introducing new students to security~\cite{ford2017capture}, formative assessment~\cite{Chothia2015a}, and as part of teaching in online universities~\cite{chicone2018using}. This spread of application areas demonstrate the use of these exercises within education. 

\section{SherLOCKED: A Detective-themed Serious Game for Security Education} 
\label{sec:game}
\subsection{Game Context} 

SherLOCKED aims to provide a game-based platform to support students in learning about cyber security. We targeted our game at consolidating content presented in undergraduate lectures. Therefore students could attend lectures and then play the game to check and refresh their understanding on certain topics. The game was conceived and deployed during the COVID-19 pandemic specifically to help address the challenges of low student motivation and poor cognitive engagement with online teaching~\cite{aguilera2020college}. This was a notable issue based on our own internal  student consultations where there was a significant difficulty in finding the motivation to watch lecture recordings, attend live online sessions, or study more generally. 

A series of in-game questions form the basis of SherLOCKED and dictate how players progress. These questions have been created using lecture content from the first half of an introductory cyber security module at the University of Kent, UK. This focus was motivated by the fact that the concepts included in these lectures form the building blocks for the remainder of the module. 

There are three levels, each containing topics aligned with the lectures of the module. For the first level, the focus is placed on the Confidentiality, Integrity and Availability (CIA) triad, and understanding the meaning of key  cyber security terms. The questions for the second level are centred on security attacks and their types, and related security services. The third level broadens the topic base and poses questions about the activities within the security risk management life-cycle (i.e., identify, analyse, treat and monitor). This allows students to learn more about cyber risk and the ways it can be managed and mitigated.

\subsection{The Game} 
SherLOCKED is a multi-level, top-down 2D detective-themed game which involves the player assuming the role of a detective navigating through a house. The player controls the movement of the detective (using arrow keys or WASD) and, as they move around each level, they have to find the questions attached to objects in each room and answer them. The game was designed using various gamification principles~\cite{fortes2018theory} and built using the Unity game engine\footnote{https://unity.com/solutions/game}---a decision made especially due to its cross-platform game deployment capabilities. Below we explain our motivation in support of key design decisions.  
\subsubsection{Theme}
A primary decision in designing a game is the theme on which to base it. Reflecting on current literature, we found that role-playing games often performed well and were preferred by students~\cite{mostafa2019development,scholefield2019gamification}. We then considered various types of roles and settled on a detective theme with an animated detective as the player's character and three cases to solve, based on the three levels (and areas of content) identified earlier. This theme was also, in part, motivated by discovery and detective games, and had a retro interface given the popularity of such games of late~\cite{heubl2020retro}.

\subsubsection{Narrative}

\begin{figure}[!ht]
  \centering
  \begin{minipage}[b]{0.48\textwidth}
    \includegraphics[width=\textwidth]{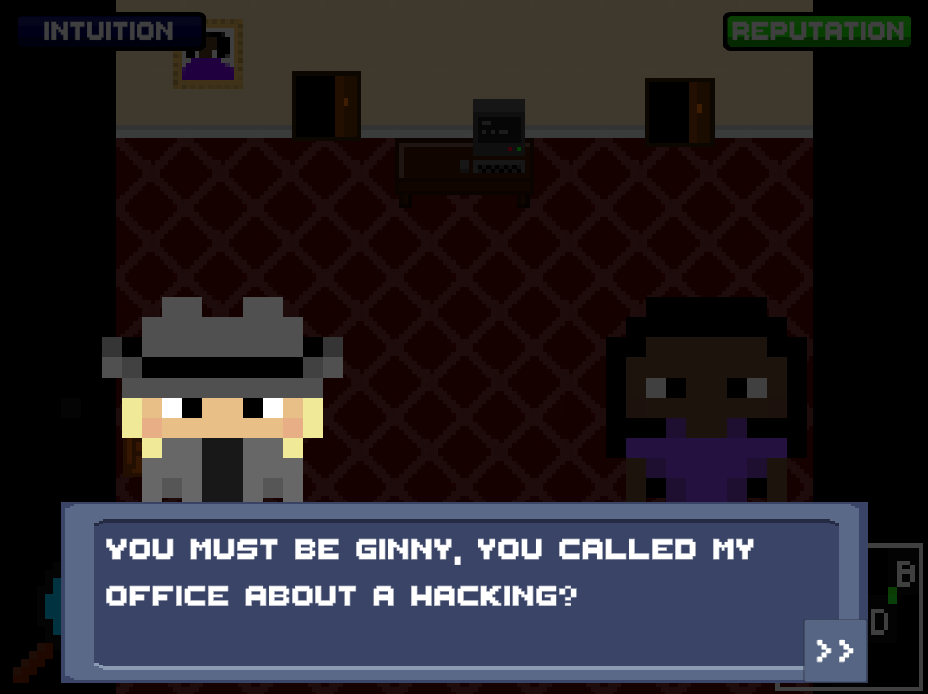}
    \caption{Introducing the player to the game's narrative. Sherry (left) is currently speaking to Ginny (right) about the cases.}
    \label{fig:1}
  \end{minipage}
  \hfill
  \begin{minipage}[b]{0.48\textwidth}
    \includegraphics[width=\textwidth]{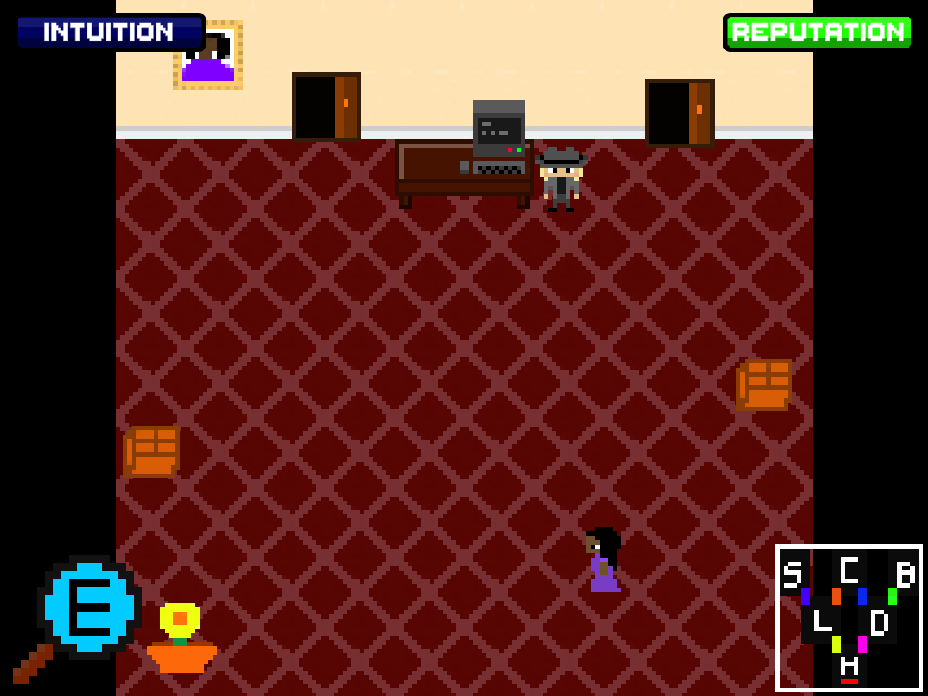}
    \caption{The game's 2D interface. The `E' in the bottom left highlights a clue has been found. A mini-map is depicted in the bottom right.}
    \label{fig:2}
  \end{minipage}
\end{figure}

A strong and compelling narrative can help players become immersed in the game, which can in turn support their learning. Gamification principles point to the key value of narratives in providing meaning to actions and in building a player's commitment to the game's tasks~\cite{fortes2018theory}. SherLOCKED's plot follows a detective named Sherry, the player character, helping the victim of a hacking, named Ginny. Figures~\ref{fig:1} and \ref{fig:2} show the opening two screens.

The narrative is told through the questions asked in the game. In the first case, the detective meets the victim and will go through the hacked home looking for clues. A clue is a question about cyber security that the player (Sherry) must answer to prove to Ginny that she's up to the task. As Sherry traverses the house, there are also various locked rooms, each room represents a new case that can only be accessed by answering enough questions correctly on the current level. 

In the second case, Sherry is in the computer room. Here, she tries to learn more about the hacker's targeting of Ginny, and as such the questions are based around cyber-attacks and how they might compromise security services. The third case involves the detective walking around the various rooms with the goal of securing the house from the risk of future hacking attacks; questions therefore are about cyber risk and high-level risk treatment solutions.

\subsubsection{Feedback}
Similar to a traditional educational context, feedback within serious games is critical in supporting learning. Feedback allows students to feel responsible for achievements, and to gauge their progress towards the preset goal (e.g. solving the case or learning about certain security topics)~\cite{fortes2018theory}. We implemented this in various ways across SherLOCKED. For instance, when a player answers a question they are always provided with informative feedback, and we provide wider progress monitoring feedback as well, as outlined later.

\begin{figure}[!h]
  \centering
  \begin{minipage}[b]{0.48\textwidth}
    \includegraphics[width=\textwidth]{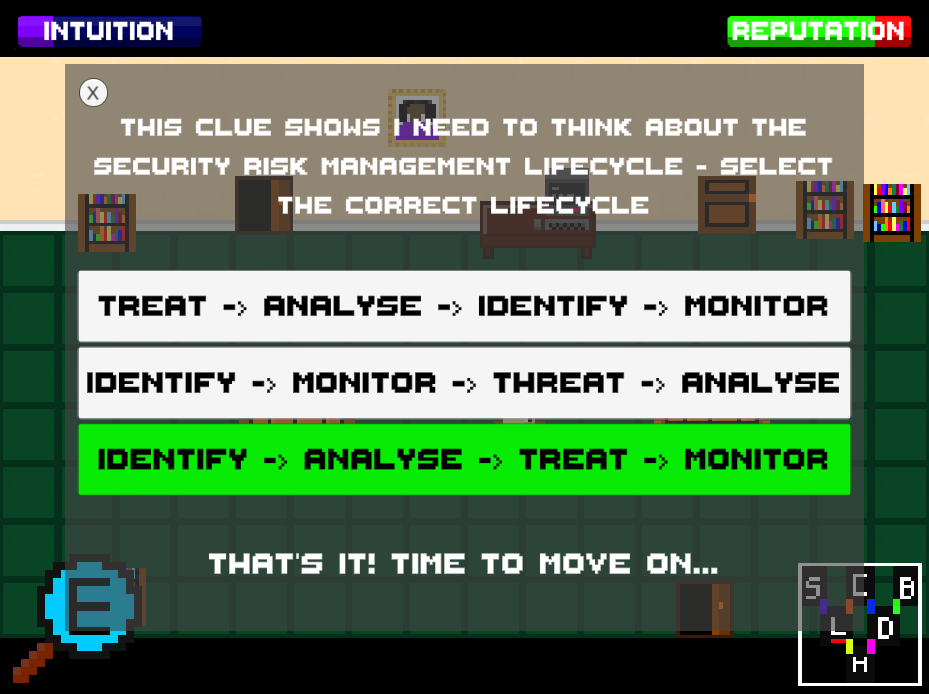}
    \caption{Correct answer with praise and Intuition points given}
    \label{fig:3}
  \end{minipage}
  \hfill
  \begin{minipage}[b]{0.48\textwidth}
    \includegraphics[width=\textwidth]{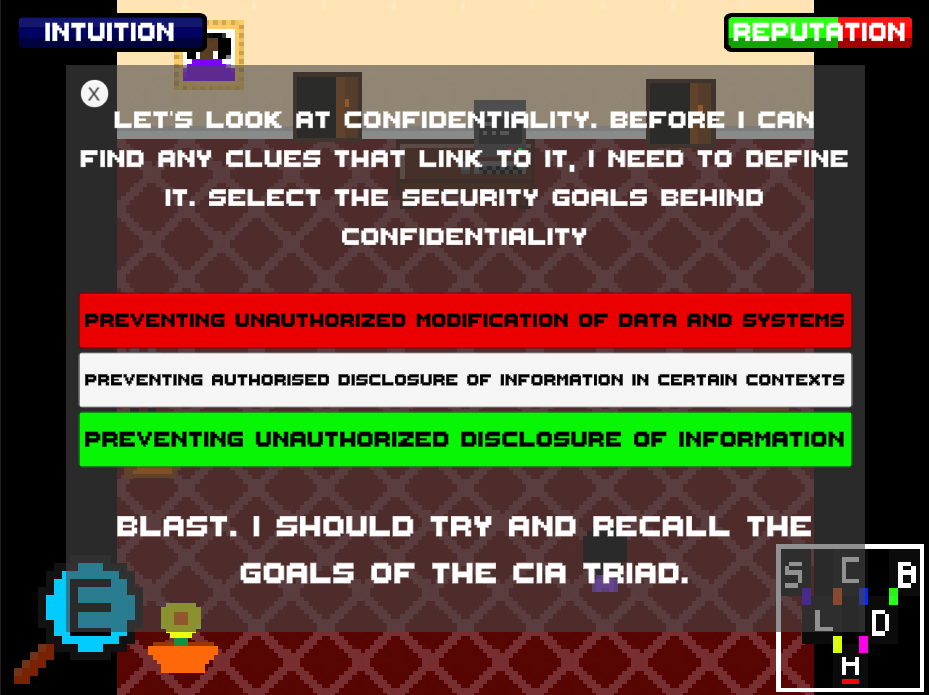}
    \caption{Incorrect answer with feedback and Reputation points lost}
    \label{fig:4}
  \end{minipage}
\end{figure}

More specifically, if players answer a question correctly, they are informed as such and the answer button they selected turns green (Figure~\ref{fig:3}). If they answer incorrectly, the answer button they selected will turn red, the correct answer button will become green, and they will be given feedback on their selection (Figure~\ref{fig:4}). This answer feedback is written to be constructive to aid learning. One study found that positive feedback was more effective than negative feedback when repair could not be made~\cite{welbers2019gamification}. Questions can only be answered once, so positive feedback is given when a question is answered incorrectly. In Figure~\ref{fig:4} for instance, we gently nudge the player to reflect on the CIA triad---given their answer suggests that the student has mixed up some concepts---before playing the level again.

\subsubsection{Progress Reporting}
Studies have found that progress reporting is a powerful gamification principle to motivate students~\cite{barata2013engaging,fortes2018theory}. The game implements two progress-typed bars to support this, and the use of levels implemented via different cases. The intuition bar represents how good the player is at picking up on clues and the reputation bar represents the player's reputation as a detective. At the beginning of each level, the intuition bar is empty and the reputation bar is full. When a question is answered incorrectly, the player (Sherry's role) loses reputation experience points (XPs) and when a question is answered correctly, they gain intuition XPs (see top left and right in Figure~\ref{fig:3} and Figure~\ref{fig:4}). 

\begin{figure}[!ht]
  \centering
  \begin{minipage}[b]{0.48\textwidth}
    \includegraphics[width=\textwidth]{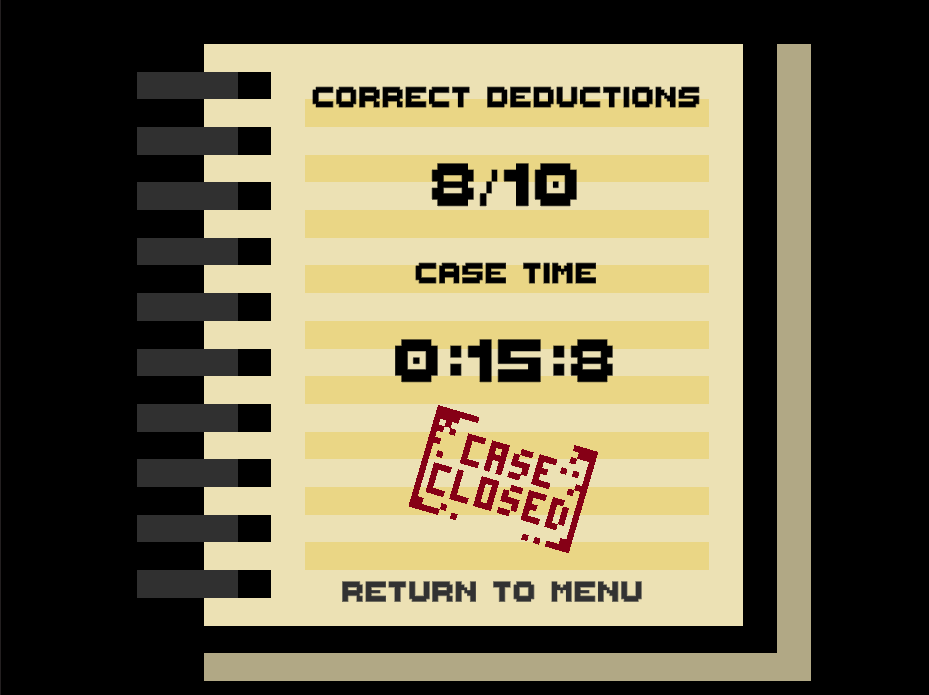}
    \caption{Case closed as the player has answered sufficient questions correctly on that level to proceed.}
    \label{fig:5}
  \end{minipage}
  \hfill
  \begin{minipage}[b]{0.48\textwidth}
    \includegraphics[width=\textwidth]{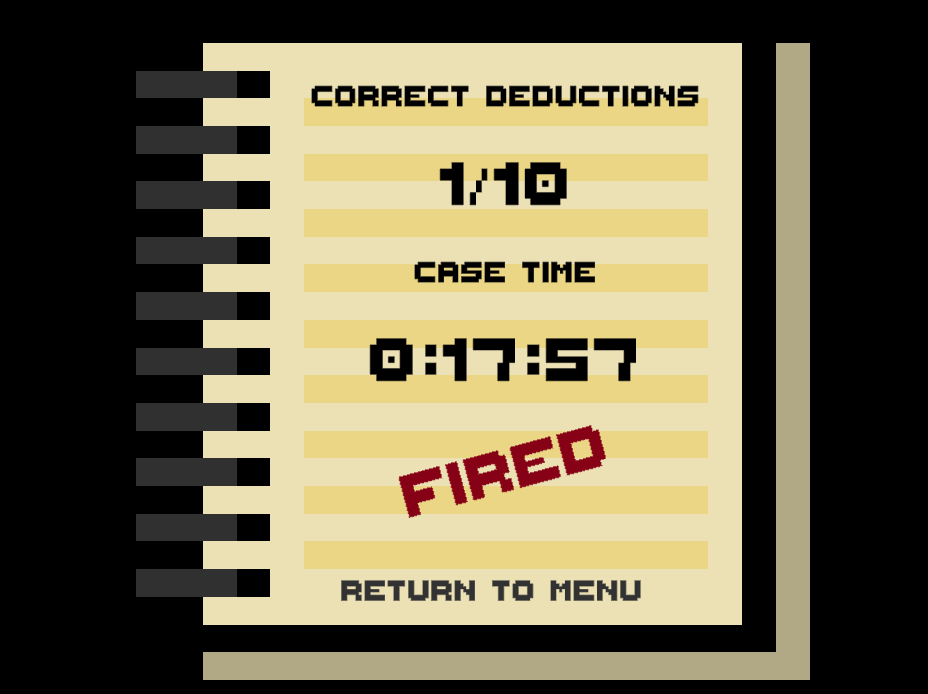}
    \caption{Player fired due to too many wrong answers on the case (and a full loss of reputation).}
    \label{fig:6}
  \end{minipage}
\end{figure}

If all intuition XPs are gained, the player is shown their score and a notification that they have ``closed'' the case on that level (Figure~\ref{fig:5}). This presents an achievement within the game and allows them to progress. If all reputation points are lost, the level is over and the player is shown their score and a notification they have been ``fired'' by Ginny from the case (Figure~\ref{fig:6}). 

\subsubsection{Competition}
Competition allows players to compare their achievements and progress against others. A common way that games implement this is using leader boards~\cite{fortes2018theory,marczewski2020table}. In SherLOCKED, when a player finishes a level, they are presented with a score value (number of questions they answered correctly) and the time taken to complete a level (e.g. Figure~\ref{fig:5}). This allows for the creation of leader boards and for players to post and compare scores, which creates competition; driving players to achieve better scores within the game and improving their learning of the covered topics.

\section{User Study and Discussion} 
\label{sec:eval}

\subsection{Study Design}
To investigate the use of SherLOCKED at supporting student learning, we conducted a user study. The study took the form of an anonymous online survey which was disseminated to undergraduate students within an introduction to cyber security module, offered to second and final year students, at the University of Kent, UK. The study received ethical approval by the university and informed consent was sought from all participants before completing the survey.  

As the game covered the first series of lectures, the study was conducted two weeks after the last of that content was delivered. Students attending the online teaching session where we launched the survey were briefed about the purpose of the game, the study and what we were aiming to achieve. They were also informed that the game would stay open for the remainder of the term if they wanted to use it to help prepare for exams. 

Before playing the game, we asked students about how confident they were in their understanding of the lecture material (using a 5-point Likert scale, with 1 as not very confident and 5 as very confident), their preferred learning methods, and what features they thought were most important when playing a game. We then provided them with an online link to the game and allowed approximately 30 minutes of play time. After they had played the game, they were asked to conduct a post-test survey which again asked the confidence questions but also requested further feedback on the game and their experience. 

\subsection{Results and Discussion}
A total of 112 students completed the survey, out of a potential 198 that were in the live lecture support session. While low (57\%), this was already encouraging given response rates for such optional, anonymous university surveys is often 20-30\% (pre-COVID-19). Considering that a main aim of the game was to support learning, we first assessed their perceptions on whether the game had an impact on how much they understood about lecture material and how confident they were about syllabus content after playing it. 

\begin{figure}[!ht]
  \centering
    \includegraphics[width=.7\textwidth]{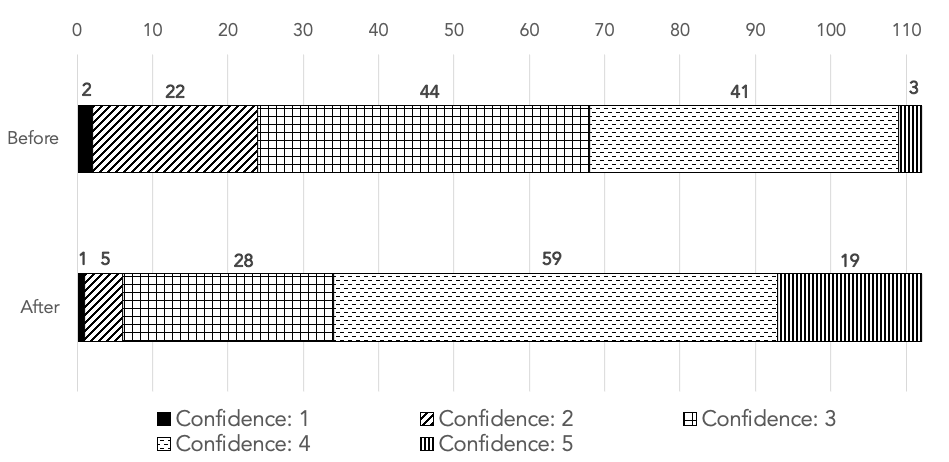}
    \caption{Confidence levels of students before and after playing SherLOCKED. This presents results on a 5-point Likert scale, with 1 as not very confident and 5 as very confident.}
    \label{fig:7}
\end{figure}

We found that 87.5\% (98) of students felt that the game helped improve their understanding of lecture content and as depicted in Figure~\ref{fig:7}, student's confidence levels increased after they played SherLOCKED. It was also encouraging to find that 65.2\% (73) stated that they would return to SherLOCKED to help revise for the exam and more generally, that 84.8\% (95) saw playing serious games as a good way to help them to learn. These positive points align well with other existing research about the current and future value of gamification when applied to higher education learning~\cite{mostafa2019development}. 

When asked about their most liked part of the game, students reported enjoying its simplicity and `retro’ feel. One participant stated the game was ``more engaging than reading notes'' and that it ``reinforced knowledge that I had, [and] helped to show areas that I was unsure about''. Another commented that it was particularly ``helpful where some of the definitions are similar and makes you think about which is the right one''. This highlights the importance of the game to be interesting and simple but still focused on adequately engaging students with content. To further note the fun nature of the game, we had included a small number of obviously incorrect answers. A participant picked up on this expressing, ``I like how sometimes the answer is obvious as other options are hilarious. It helps me when I’m unsure about the answer and acts as a revision''. Again, this is a useful point for future research in this domain as it ensures that games keep a good balance between being serious and enjoyable. 

Students also valued the instantaneous feedback after answering questions. As was hinted at in the quotes above, these helped direct them to relevant materials and to discover areas where their understanding or knowledge may be lacking. We also noted the significant importance of the game's theme and narrative to students. To summarise with one student's comment, ``The setting fits the theme of the game. It makes sense to teach cyber sec concepts in a detective setting so it doesn't feel completely out of place like some learning games''. A key takeaway point therefore is that serious games may be better tailored to the subject context instead of being generic. 

Some other general findings were of note as well. Firstly, students reported that their preferred learning styles were primarily attending lectures (33\% of respondents, i.e., 37 students) and engaging in question-answer sessions (46\%, 52). Reading notes and textbooks (11\%, 13) or writing notes (9\%, 10) were not favoured. For serious games in university contexts therefore, this may be an opportunity to pair games (or levels) with individual (or groups of) lectures. We also found that when playing a game, if students had to choose one feature to represent the most important feature, feedback on actions (27\% of respondents, i.e., 30 students), competition (24\%, 27) and narrative (23\%, 26) were key. Less so were character customisation (2\%, 2), a consistent theme (e.g. RPG) (4\%, 5), an informative tutorial (8\%, 9) and progress reporting (12\%, 13). While all of these are ideal in a game and we would certainly encourage educators to strive for them, this can help prioritise features/principles if time or resources are low.

There were some areas for improvement in our work that were identified by participants. This arose through a query where we asked what they liked least about the game. The main theme which emerged from a few participants was about the simplicity of the game; a notable point given its simplicity was praised by others. Here, a participant stated ``the gameplay was quite boring, being just walk to object, interact, answer quiz, go to next object, with not much reward (apart from not losing)''.  This links to another participant comments on how it ``would be cool with gold coins to collect''. These are clearly valid points and ones that could make the game more engaging. We are currently exploring the feasibility of different interface designs (e.g. integrating the ability to track a hacker), different style questions (e.g. a puzzle instead of multiple choice), and how badges or coins may be used (e.g. a badge for fastest time, or receiving in-game Bitcoin for successfully finding out information about the hacker). 

\section{Conclusion and Future Work} 
\label{sec:con}

While serious games have been discussed for decades, their uptake in higher education seems to be limited. These games however can offer a great deal to a student's learning experience. Considering this value, this article proposed SherLOCKED, a new serious game created in the style of a 2D top-down detective adventure. SherLOCKED was designed to be used within an undergraduate cyber security course to complement lecture materials and content. A key motivation of the game was to overcome the challenges that students were facing with online-only learning because of the COVID-19 pandemic, including a lack of motivation and cognitive engagement with materials. 

From our user study with a cohort of 112 university students, we found that the game was successful at providing a more enticing form of interaction which also assisted student's understanding and confidence with lecture content. We were also able to identify a number of key take away points for research in this domain more generally, including the importance of a strong narrative, the balance of informative and fun interfaces, game stories that are tailored to the subject context, and actionable and immediate feedback. 

There are various avenues for future work, but there are two of particular interest. The first is to extend upon the game concept currently built. Currently the game only covers the initial lectures in the module, and therefore the others will need to be designed and implemented. Additionally, as mentioned above we would aim to explore ways to make the game more engaging and avoid issues of boredom (e.g. badges, interfaces and question styles). The second goal is to setup a long-term study to explore whether the expanded game results in better actual ongoing student engagement with content and actual student performance. This aims to tackle the current reliance on self-report data which can be biased. We believe that while the current version of SherLOCKED performed well, with improvements in the interface and via an extended user study, the game will be one that can be used within the undergraduate module for years to come. 

\bibliographystyle{splncs04}
\bibliography{main}
\end{document}